\documentclass{optica-article}
\journal{opticajournal} 

\usepackage{siunitx}				
\usepackage{textcomp}               

\usepackage{booktabs}               

\newcommand{\contrastenhancement}{\ensuremath{\eta_\sigma}}

\begin{document}

\title{Model-based aberration corrected microscopy inside a glass tube}

\author{D.W.S. Cox\authormark{1, *}, T. Knop\authormark{1} and I.M. Vellekoop\authormark{1}}

\address{\authormark{1}Biomedical Photonic Imaging Group, Faculty of Science and Technology, University of Twente, P.O. Box 217, 7500 AE Enschede, The Netherlands}

\email{\authormark{*}Corresponding author: d.w.s.cox@utwente.nl}

\begin{abstract*}
Microscope objectives achieve near diffraction-limited performance only when used under the conditions they are designed for. In non-standard geometries, such as thick cover slips or curved surfaces, severe aberrations arise, inevitably impairing high-resolution imaging. Correcting such large aberrations using standard adaptive optics can be challenging: existing solutions are either not suited for strong aberrations, or require extensive feedback measurements, consequently taking a significant portion of the photon budget. 
We demonstrate that it is possible to pre-compute the corrections needed for high-resolution imaging inside a glass tube based on \emph{a priori} information only. Our ray-tracing based method achieved over an order of magnitude increase in image contrast without the need for a feedback signal. 
\end{abstract*}

\section{Introduction}
Microscope objectives are designed to work with a specific immersion medium and, optionally, a cover slip. In the correct geometry, microscope objectives allow the study of specimens at a resolution that is close to the diffraction limit. However, in some applications the use of a non-standard geometry with thick cover slips, mismatching refractive index, or even curved surfaces is unavoidable. For example, lumen-based organ-on-a-chip devices aim to model tubular organs. The tubular shape in these devices is a key aspect to recreating organ-level structure-function relationships as observed in vivo \cite{virumbrales-munozMicrofluidicLumenbasedSystems2020, bischelImportanceBeingLumen2014, lowOrgansonchipsNextDecade2021}.
Since even a small refractive index mismatch causes severe aberrations, especially for high-NA objectives \cite{schwertnerCharacterizingSpecimenInduced2004}, high-resolution imaging inside such 3-D structures is practically impossible. 

\begin{figure}[b]
    \centering
    \includegraphics[width=0.8\linewidth]{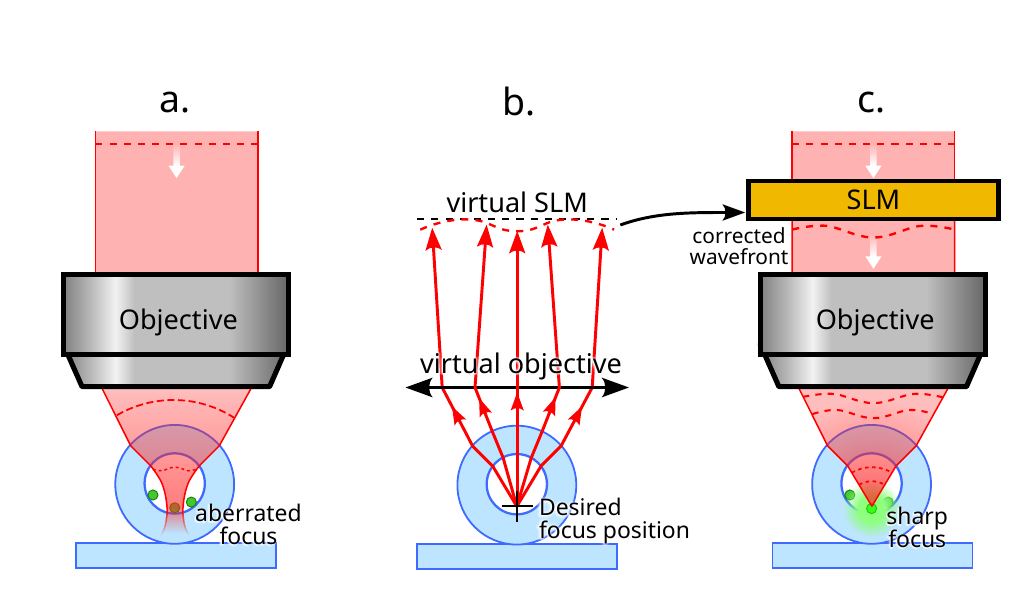}
    \caption{
    a. Without any correction, focusing inside a tube causes an aberrated focus. This prevents high resolution imaging.
    b. a model is used to compute a phase pattern to correct the focus at a specified location.
    c. A spatial light modulator is used to display the computed corrections. A sharp focus is formed, enabling high resolution imaging.}
    \label{fig:raytrace-method}
\end{figure}
In principle, adaptive optics (AO) can be used to correct for aberrations. In AO, a deformable mirror (DM) or spatial light modulator (SLM), creates a wavefront that specifically counteracts the aberrations induced by the sample \cite{boothAdaptiveOpticsMicroscopy2007, kubbyAdaptiveOpticsBiological2013}. However, existing methods are either ill-suited for the extreme aberrations found in these structures, or require extensive feedback measurements, consequently using a significant portion of the photon budget.

AO often relies on the presence of a point-like fluorescent particle, known as a `guide star'. Using a wavefront sensor or interferometric techniques, the wavefront of the light emitted by the guide star can be measured, and phase-conjugated to compensate the aberrations\cite{taoAdaptiveOpticsConfocal2011,horstmeyerGuidestarassistedWavefrontshapingMethods2015, vellekoopDigitalOpticalPhase2012}. Alternatively, the image quality or signal strength from the guide star can be optimized using an algorithm that iteratively modulates the wavefront to maximize a feedback signal, such that a sharp focus is formed. These methods are known as sensorless AO \cite{debarreImagebasedAdaptiveOptics2009}.
For strong and complex aberrations, finding the correction typically requires a large number of measurements. This not only takes time, but extensive illumination can also result in phototoxicity and/or photobleaching of the fluorophores \cite{boothAdaptiveOpticsMicroscopy2007}. Furthermore, for many of the more efficient AO algorithms, convergence to the global optimum is not guaranteed when the wavefront error exceeds several radians \cite{siemonsRobustAdaptiveOptics2021}. 

The problem of convergence is solved with the introduction of wavefront shaping \cite{vellekoopFocusingCoherentLight2007a, kubbyWavefrontShapingBiomedical2019}. Wavefront shaping algorithms are guaranteed to converge to the optimal correction for any sample \cite{vellekoopFocusingCoherentLight2007a}, and allow focusing light even through strongly multiple-scattering materials. Wavefront shaping algorithms can be optimized to cover the full range from weak aberration to strong scattering \cite{kubbyWavefrontShapingBiomedical2019, mastianiWavefrontShapingForward2022}. However, still hundreds of measurements are needed to find the optimal correction.

The use of guide stars also has practical and fundamental limitations. Firstly, it may not always be possible to incorporate point-like guide stars or achieve an acceptable signal to noise ratio (SNR) in order to form a good focus. Second, once a correction is found, it can only be used within a small region around the guide star, known as the isoplanatic patch \cite{kubbyWavefrontShapingBiomedical2019}.

Recently, a third class of methods has started to emerge, which we will refer to as \emph{model-based}. These methods rely on the existence of an accurate 3-D computer model of the sample, a \emph{digital twin}, which is used to simulate a wavefront sensing experiment with a perfect point-source. This way, the correction pattern can be computed and used directly, without the need for feedback measurements, guide stars or wavefront sensors. 

A major advantage of the model-based approach is that the same model can be used to obtain wavefront correction patterns for arbitrary positions inside the sample, thereby eliminating the restriction imposed by a finite isoplanatic patch.

The model-based approach was demonstrated for flat interfaces\cite{salterExploringDepthRange2014, matsumotoCorrectionDepthinducedSpherical2015,turagaAberrationsTheirCorrection2013, matsumotoAberrationCorrectionConsidering2018}, for several smooth surface shapes \cite{yamaguchiAdaptiveOpticalTwoPhoton2021a} and for a thin scattering layer \cite{thendiyammalModelbasedWavefrontShaping2020}. 
For the latter two, imaging of the sample surface is required in order to create an accurate 3-D model. 

In this work, we demonstrate the use of model-based wavefront corrections for imaging inside large cylindrical structures. We show that it is possible to compute an accurate correction pattern based on \emph{a priori} information only, and that the correction is on par with an exhaustive-search AO algorithm. We systematically analyse the method's sensitivity to its model parameters and show that \emph{a priori} knowledge of these parameters is sufficient to produce an accurate phase correction pattern.

Our method is illustrated in Fig. \ref{fig:raytrace-method}. We consider a geometry with fluorescent particles inside a glass tube. Without correction, refraction at the surfaces of the tube causes severe aberrations, impairing the formation of a sharp focus (Fig. \ref{fig:raytrace-method}a). Before starting the imaging experiment, we model the tube and microscope objective, and use a ray tracing simulation to compute the correction pattern (Fig. \ref{fig:raytrace-method}b). We then apply the computed correction using an SLM (Fig. \ref{fig:raytrace-method}c) and demonstrate enhanced imaging inside a glass tube, with an order of magnitude signal intensity increase compared to imaging without correction.

\section{Method}
We demonstrate our method in a laser scanning 2-photon excitation fluorescence (2PEF) microscope, with a spatial light modulator conjugated to the back pupil plane of a water-dipped objective, as described in \cite{thendiyammalModelbasedWavefrontShaping2020}.

The sample is a capillary tube (Micropipettes 0.5µL, Cat.nr. 1-000-0005, Drummond Scientific) filled with an \qty{4.5}{g/L} agar solution in water, mixed with fluorescent beads (Fluoresbrite\textsuperscript{\textregistered} YG Carboxylate Microspheres \qty{0.50}{\micro m}, Cat.nr. 15700, Polysciences). The tube was glued to a microscope slide (ISO\,8037-1) to make it compatible with existing microscope sample holders. With a bright field microscope, we found that the tube has an outer radius of \qty{286.5 +- 1.5}{\micro m} and shell thickness of \qty{215.5 +- 1.6}{\micro m}, where the error margins indicate standard deviation. The tube is made of BK7 borosilicate glass, with a refractive index of \num{1.5106 +- 0.0005} \cite{DatasheetsDownloadsOptical}.

\subsection{Phase correction computation}
We model the tube as two cylindrical interfaces, based on the known inner and outer radius and refractive index of the tube. We approximate the objective as a perfect Abbe sine corrected objective, since infinity-corrected objectives are designed to closely adhere to the Abbe sine condition.
Once the 3-D model is defined, we place a virtual point source at the desired focus position inside the tube model and perform a ray tracing simulation that keeps track of the optical path length (OPL) (see Figure \ref{fig:raytrace-method}b). Using a vectorial Snell's law to compute refraction \cite{tkaczykVectorialLawsRefraction2012}, the rays are traced through the tube's interfaces. Tracing the rays further, through the objective and onto the virtual SLM plane yields the optical path length $\mathrm{OPL}(x, y)$ for each ray position $(x, y)$ on the virtual SLM. We then compute the raw phase correction $\phi(x, y)$ at SLM position $(x, y)$ as
\begin{equation}
    \label{eq:phase-OPL}
    \phi(x, y) = -k_0\mathrm{OPL}(x, y)
\end{equation}
where $k_0=2\pi/\lambda$, with $\lambda$ the wavelength of the light in vacuum.

At this point, we have computed the phase correction at each ray position. Finally, we compute the phase correction at each SLM pixel by linearly interpolating the optical path length between the simulated ray positions. We used \num{250 000} rays to compute our correction patterns. With our current implementation, this takes a few seconds per pattern on an  Intel Core i7-8700 PC. Our ray-tracing code is freely available as open source software \cite{IvoVellekoopRaylearn}.

The raw phase correction computed this way typically contains a large defocus component, resulting in a steep phase gradient which can degrade the performance of the SLM \cite{salterExploringDepthRange2014}. Since the defocus component can also be resolved by simply moving the objective along the optical axis with respect to the sample, we remove this component from the phase correction pattern. Concretely, we minimize the mean square sum of of the phase gradients by translating the virtual objective in our simulation, using the AMSGrad algorithm \cite{reddiConvergenceAdam2018} to find the optimum.

Similarly, the phase correction pattern can contain a significant tilt component. This tilt component simply causes a small translation of the image and does not affect the image quality, but it can degrade the performance of the SLM. Hence, we remove the tilt by computing the average tilt component of the simulated rays, and subtracting it from the optical pathlength.

\subsection{Sensitivity analysis}

\begin{figure}
    \centering
    \includegraphics[width=0.4\linewidth]{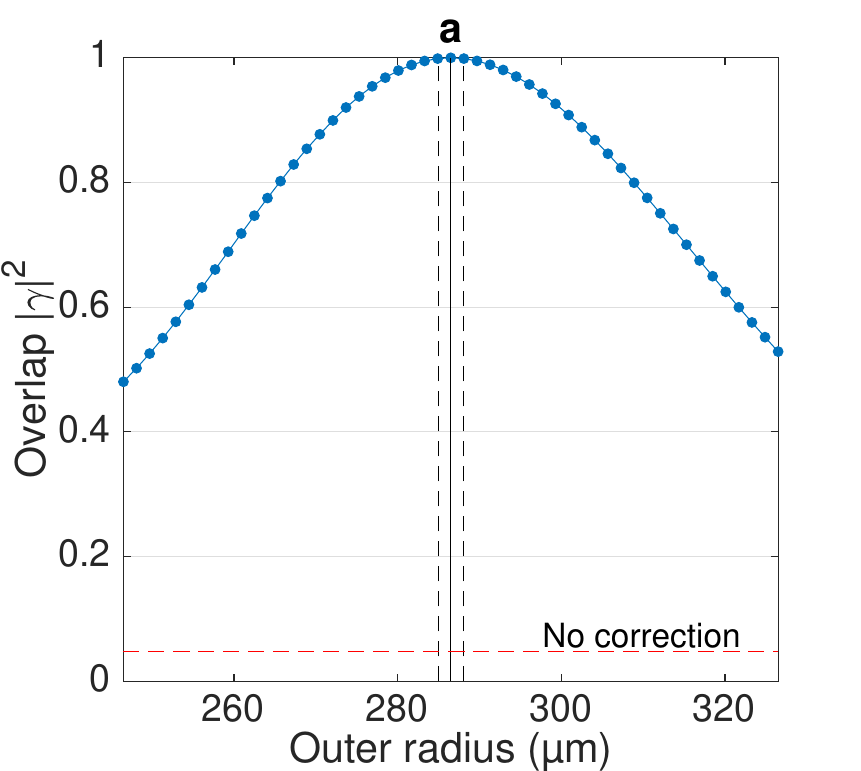}
    \includegraphics[width=0.4\linewidth]{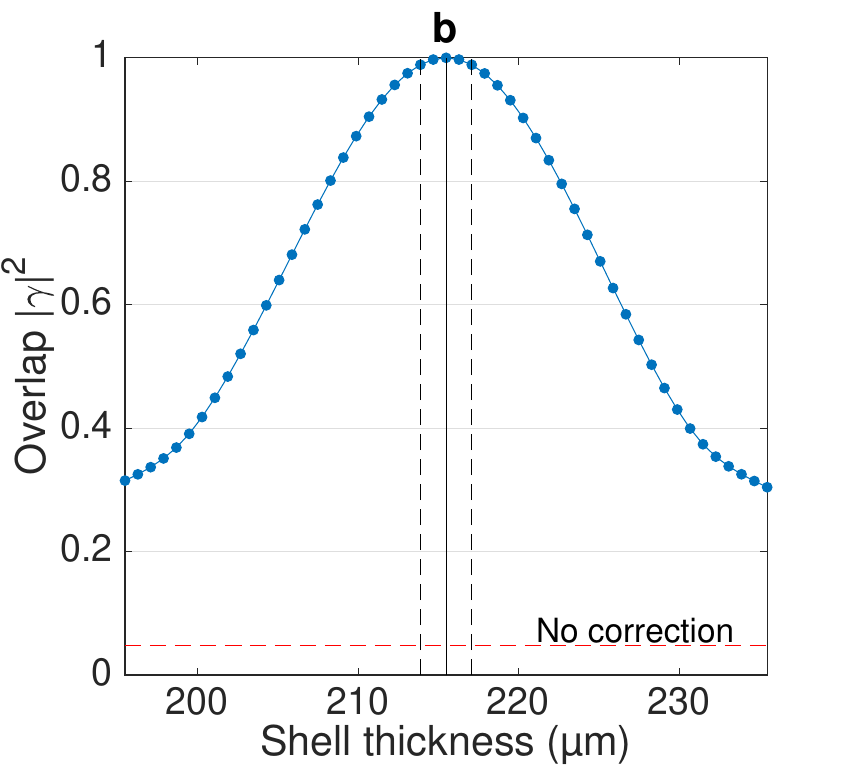}
    \\\vspace{0.5em}
    \includegraphics[width=0.4\linewidth]{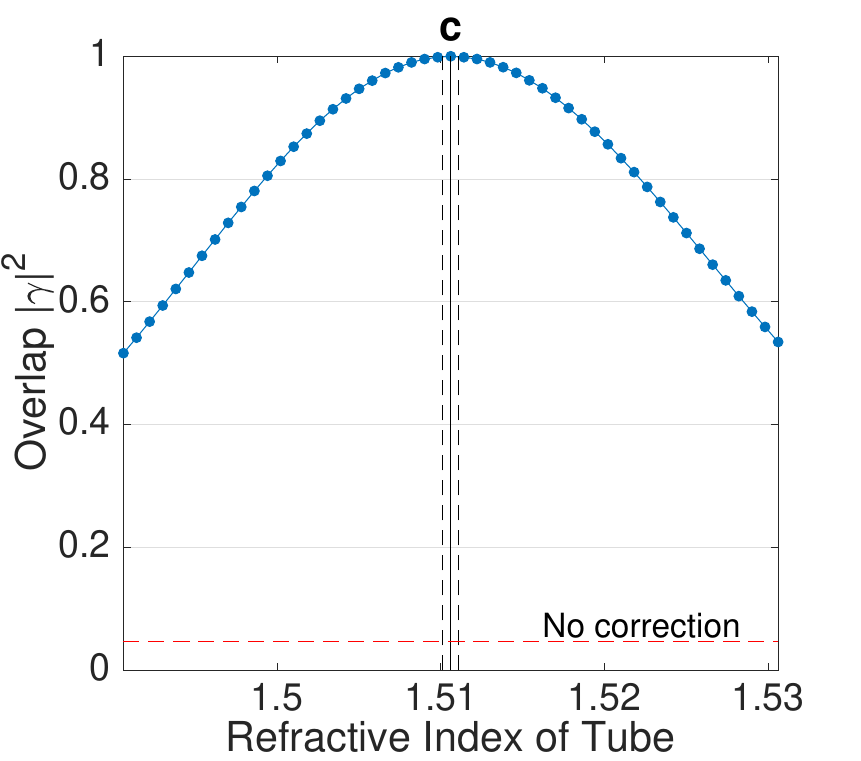}
    \includegraphics[width=0.4\linewidth]{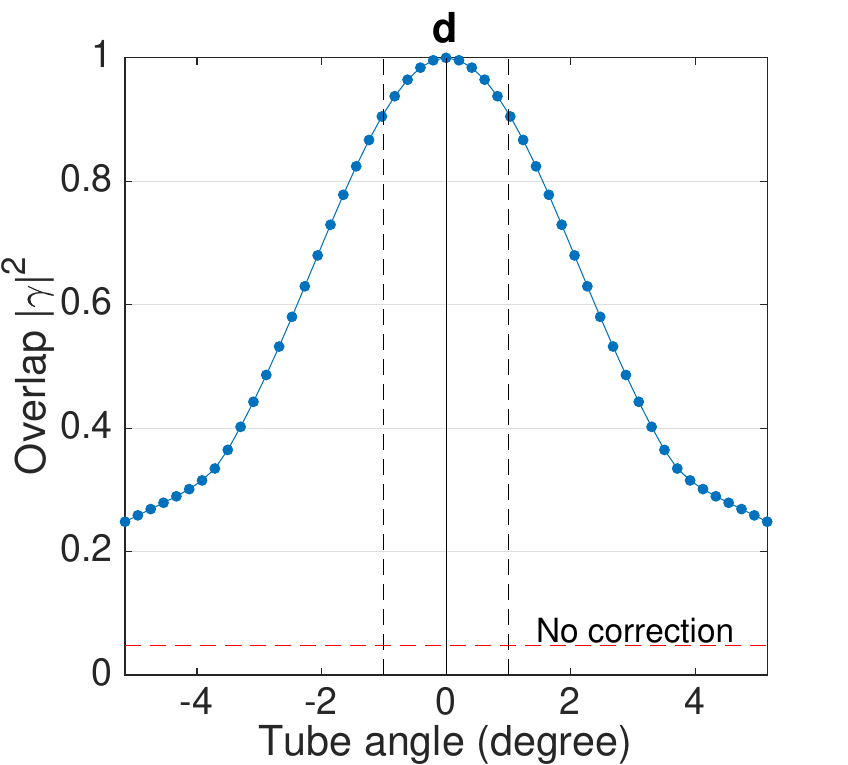}
    \caption{Sensitivity of the computed correction pattern for the bottom of the tube to deviations in the model parameters.
    For each pattern, the absolute squared overlap coefficient $|\gamma|^2$ with the unperturbed pattern is plotted as a blue line with dots.
    In each plot, solid black vertical lines indicate the unperturbed parameters. Dashed black vertical lines indicate the error margins.
    Lastly, for a flat phase pattern (i.e. no correction) we found $|\gamma|^2=0.047$. This value is indicated with red dashed horizontal lines.
    a. Sensitivity to outer radius.  b. Sensitivity to shell thickness. c. Sensitivity to refractive index of tube. d. Sensitivity to tube rotation around the optical axis.
    }
    \label{fig:parameter-sensitivity}
\end{figure}

We assess the method's sensitivity to the accuracy of the model parameters. To quantify the sensitivity, we use the degree of intensity control $|\gamma|^2$, where $\gamma$ is the overlap coefficient as defined in Equation 3 of  \cite{vellekoopUniversalOptimalTransmission2008}: 
\begin{equation}
    \label{eq:overlap-coefficient}
    \gamma \equiv \sum_{x, y}\left(E_1\right)^* E_2
\end{equation}

where $x, y$ denote all contributing SLM pixels. $E_1=A e^{i\phi_1}$, with $\phi_1$ the phase pattern to correct a focus close to the bottom of the tube, computed with the aforementioned values for the parameters. $E_2=A e^{i(\phi_2 + \phi_d)}$, with $\phi_2$ the phase pattern for the perturbed parameter. 
As field amplitude $A$, we used a Gaussian beam profile with a waist of $1.18$ times the pupil radius, based on the measured beam profile at the SLM surface. $A$ is real and normalized such that $\sum_{x,y}A^2 = 1$. Lastly, $\phi_d$ is the defocus phase pattern that maximizes $|\gamma|^2$:
\begin{equation}
    \label{eq:defocus}    
    \phi_d = z_d k_z = z_d k_0 \sqrt{n^2 - (r\, \mathrm{NA} /R)^2}
\end{equation}

where $z_d$ is the defocus distance that maximizes $|\gamma|^2$. $k_z$ is the wavevector component along the optical axis, $n$ is the refractive index of the medium of the focus, NA denotes the numerical aperture of the objective, $r$ is the distance to the optical axis of the corresponding SLM pixel, and $R$ is the radius of the objective pupil, conjugated to the SLM plane.
We find $z_d$ by trying a small range of values and use the one that maximizes $|\gamma|^2$.
This optimization compensates for possible mismatching defocus in the patterns.

Note that $|\gamma|^2$ corresponds to the fraction of the incident energy that is present in the perfectly shaped mode, where the rest of the energy is distributed over other modes that do not contribute to the focus \cite{vellekoopUniversalOptimalTransmission2008}. In this case, therefore, $|\gamma|^2$ is equivalent to the Strehl ratio (Eq. 4.11 from \cite{kubbyAdaptiveOpticsBiological2013}).

In Figure \ref{fig:parameter-sensitivity}, we show $|\gamma|^2$ as a function of the model parameters. The vertical solid black lines indicate the model parameters used to compute the correction, and the dashed lines indicate the accuracy to which these parameters are known. Importantly, for a flat pattern (i.e. no correction), a value of only  $|\gamma|^2$= \num{0.047} was found. Although exact quantification of signal enhancement is difficult for fluorescent beads \cite{sinefeldAdaptiveOpticsMultiphoton2015}, we can certainly expect a significantly improved contrast when correcting the aberrations.

We observe that the overlap coefficients of the outer radius (Figure \ref{fig:parameter-sensitivity}a), shell thickness (Figure \ref{fig:parameter-sensitivity}b) and refractive index (Figure \ref{fig:parameter-sensitivity}c) are above $|\gamma|^2 > 0.98$ within the error margins. Hence, we conclude that these parameters are sufficiently accurate to compute the correction pattern.

As can be seen in Fig. \ref{fig:parameter-sensitivity}d, the correction pattern is the most sensitive to the orientation of the tube. For this reason, we rotated the pattern on the SLM to be aligned with the orientation of the tube before performing the experiments. This alignment was done by  rotating the pattern over a small range of angles to maximize the fluorescence signal. We estimate this calibration reduces the tube angle error to approximately \ang{0.1}, which makes it sufficiently accurate. After this one-time alignment, our method allows us to correct the focus anywhere in the tube, without any additional feedback measurements.

\section{Results}

\begin{figure}
    \centering
    \includegraphics[width=1\linewidth]{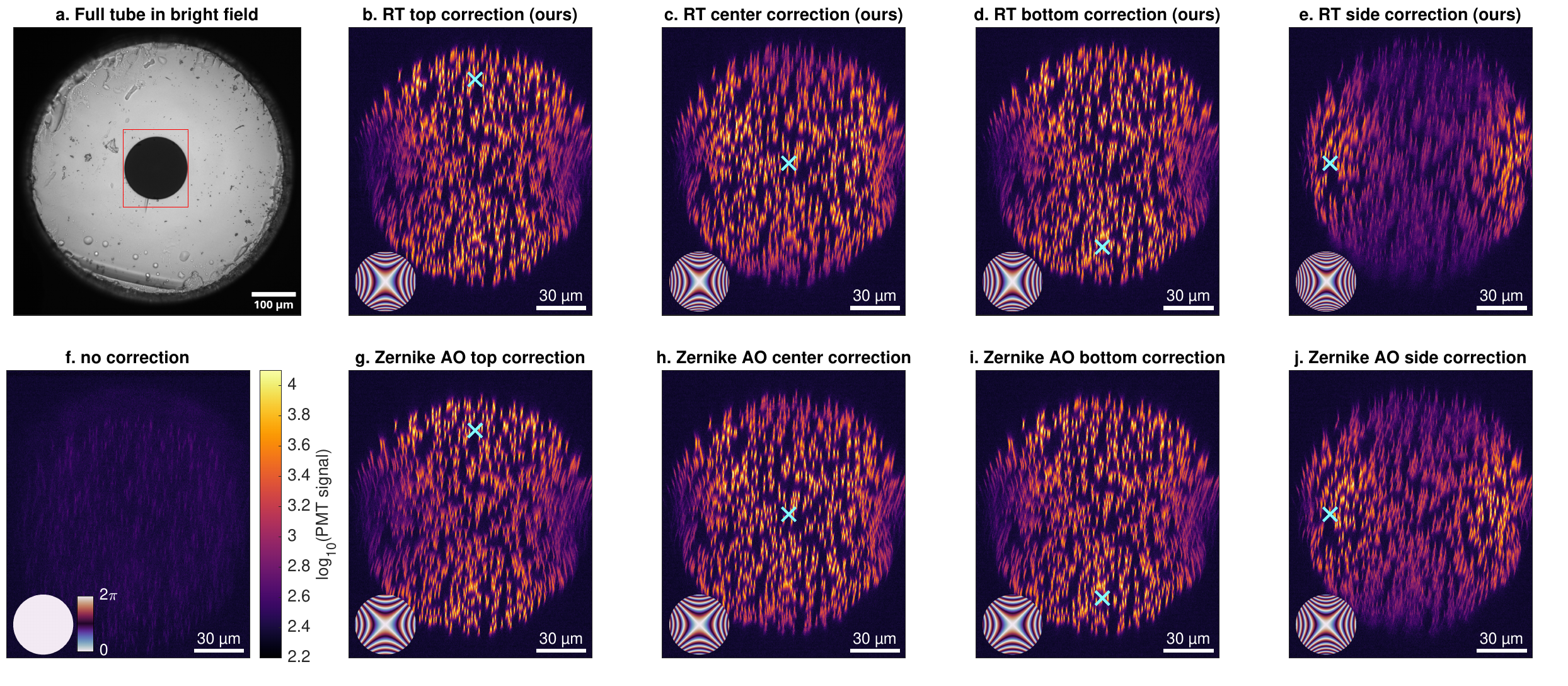}
    \caption{a: Bright field image of the capillary glass tube. The tube has an inner diameter of \qty{142}{\micro m} and an outer diameter of \qty{573}{\micro m}. The red rectangle indicates the volume imaged with 2PEF. b-j: 2PEF images of \qty{0.5}{\micro m} fluorescent beads in agar inside a capillary glass tube. All fluorescence images are 99.5-percentile projections of the imaged volume. The color bar indicates the $\log_{10}$ of the signal intensity.
    f. Without aberration correction.
    b-e, g-j: Aberration corrected images. The circular tube inner cross-section is clearly visible.
    Each fluorescence image includes an inset with the corresponding phase correction pattern that was used. Since the optical pathlength difference within each correction pattern greatly exceeds the wavelength, the phase correction pattern is wrapped to $[0, 2\pi]$. The patterns were computed/optimized for various target focus locations, which are marked by a cyan $\times$.
    b-e: correction patterns computed using our ray tracing method. g-j: correction patterns acquired by scanning two Zernike modes in a feedback-based brute-force grid-search. The color scale indicates the $\log_{10}$ of the signal intensity of the photon multiplier tube, and is the same for all images.}
    \label{fig:results-tube-img}
\end{figure}

We used our methods to image fluorescent beads inside the glass tube. Figure \ref{fig:results-tube-img}a shows a bright-field image of the side of the tube, whereas the other images are taken with the 2PEF, and show show 99.5-percentile projections of the 3-D image stack of the cross section of the tube.  The circular shape of the tube's cross section is clearly visible in each image. Every 2PEF image includes an inset showing the phase correction pattern used for that image. Without any aberration correction (Figure \ref{fig:results-tube-img}f), the beads show up as faint, blurry blobs.

Figures \ref{fig:results-tube-img}b-e (in the top row labeled `RT') show the effect of applying the corrections obtained with our ray-tracing model. It is clear that every tested phase correction pattern significantly increases the signal strength, resolution, and contrast. 

The images in Fig.\ref{fig:results-tube-img}b-e are all corrected for a specific target focus location inside the tube. These locations are marked with a cyan $\times$. We observe that for each image, the signal enhancement is strongest near the target focus location. This volume is known as the isoplanatic patch \cite{kubbyWavefrontShapingBiomedical2019}.

We observe that the phase correction patterns for `Top' and `Bottom' happen to be very similar, which explains the similarity in the corrected images. We can make a similar observation about the `Side' correction. Both the left and right side of the tube display a significantly enhanced signal. At the right side of the tube, the situation is a mirrored version of the left side. Its aberrations would therefore be mirrored as well. Due to the strong symmetry in the phase correction patterns, the corresponding correction pattern significantly enhances both the left and the right side of the tube, even though it was computed to correct the left side only. Still, it is clearly beneficial to use different correction patterns for different parts of the sample, as no single correction pattern corrects the entire tube maximally. Moreover, computing these patterns requires no additional measurements and thus is essentially `free'.

\subsection{Comparison with exhaustive-search Adaptive Optics}

We now compare our model-based technique to feedback-based sensorless AO. As expected for a sample of this size, shape and refractive index mismatch, the wavefront error far exceeds the 1 radian range recommended in \cite{siemonsRobustAdaptiveOptics2021}, meaning that most algorithms used for sensorless AO are not guaranteed to converge to the correct solution. A further complication is that it is not known \emph{a priori} which modes to optimize.

In what follows, we assume a best-case scenario for sensorless AO, in which we correct for the most dominant modes and assume that the global optimum correction in this search space is found. To ensure selecting the appropriate Zernike modes, we performed a Zernike decomposition of each of our ray traced patterns, and select the two modes that comprise the majority the aberrations. We found that the main component of each of the patterns is by far primary astigmatism. With the Zernike basis as defined in \cite{brugEfficientCartesianRepresentation1997a}, we find the following primary astigmatism amplitudes `Top': \qty{30.6}{rad}, `Center': \qty{34.4}{rad}, `Bottom': \qty{31.0}{rad}, `Side': \qty{41.1}{rad}. The second largest component for the `Top', `Center' and `Bottom' locations is secondary astigmatism, with amplitudes \qtylist{2.82; 2.84; 2.81}{rad} respectively. For the `Side' location, quadrafoil is the second largest component with an amplitude of \qty{-3.82}{rad}. Hence, we optimize primary and secondary astigmatism for the `Top', `Center' and `Bottom' locations, and we optimize primary astigmatism and quadrafoil for the `Side' location.

Note that we use the same rotation calibration that we use for our ray traced patterns to align the coordinate system of the Zernike modes with the direction of the cylindrical tube. Consequently, astigmatism and quadrafoil can be fully represented by only the vertical modes, whereas the oblique modes are eliminated due to the symmetry of the tube.

To guarantee finding the global optimum in practice, we used a slow but reliable brute-force grid-search of the two dominant modes to find correction patterns for the different locations inside the tube. Each of the grid points is a measurement of the strength of the feedback signal for a specific linear combination of two Zernike modes. We use bicubic interpolation to find the optimum in between grid points.

It may be expected that sensorless AO outperforms model-based wavefront shaping because it compensates for the actual aberrations in the system rather than the aberrations in a simplified and imperfect computer model. This expected benefit comes at the cost of a lengthy optimization procedure that expends a significant portion of the photon budget. However, as shown  in the bottom row of Figure \ref{fig:results-tube-img}g-j, our model-based approach and the sensorless AO perform comparably, with visually similarly enhanced images.

To quantify the improvement of the image quality, we define the contrast enhancement metric \contrastenhancement:
\begin{equation}
    \label{eq:contrast-enhancement}
    \contrastenhancement
    = \frac{\sigma_\mathrm{signal}}{\sigma_\mathrm{reference}}
    = \frac{
        \sqrt{\sigma^2_\mathrm{corrected} - \sigma^2_\mathrm{background}}
    }{
        \sqrt{\sigma^2_\mathrm{uncorrected} - \sigma^2_\mathrm{background}}
    }
\end{equation}
where $\sigma^2$ denotes the variance of the corresponding measured image stack. To correct the contrast enhancement for the systematic effect of noise, we assume that the noise is additive and not correlated to the signal. Hence, we can subtract the variance of the noise, as taken from a background measurement with no beads present, from the image variances to get an unbiased estimate of the contrast enhancement. 

For each of the different target focus locations, and for both correction methods, we computed the contrast enhancement \contrastenhancement, as shown in the table of Table \ref{tab:contrast-enhancement}. In accordance with the corrected images of Figure \ref{fig:results-tube-img}, all correction patterns yield significant signal enhancements over uncorrected imaging. For the `Top' and `Bottom' locations, our ray tracing method performed slightly better than the sensorless AO (labeled `Zernike AO' in Table \ref{tab:contrast-enhancement}). As previously explained, our method does not express the phase correction as a finite linear combination of Zernike modes. Hence, the patterns computed by our method automatically include the higher order Zernike modes, in contrast to the limited number of Zernike modes included in sensorless AO. We speculate this could explain the slightly higher performance of our method for these locations.

\begin{table}
    \centering
    \begin{tabular}{lcccc}
        \multicolumn{5}{c}{Contrast enhancement \contrastenhancement}\\
    \toprule
        & Top& Center & Bottom & Side\\ 
    \midrule
        Model-based& 11.3& 9.0& 11.6&7.8\\
        Zernike AO& 10.5& 11.0& 10.3&11.0\\
    \bottomrule
    \end{tabular}
    \caption{Contrast enhancement {\contrastenhancement} for the tested correction patterns. The columns Top, Center, Bottom and Side correspond to the different focus locations inside the tube.
    The contrast enhancements were computed over a \qtyproduct{20 x 20 x 74}{\micro m} volume at each corresponding location.
    The row `Zernike AO' denotes the correction patterns acquired by scanning 2 selected Zernike modes in a feedback-based brute-force grid-search.
    }
    \label{tab:contrast-enhancement}
\end{table}

For the `Center' and `Side' location, our model-based method performs worse than sensorless AO. We speculate this difference could be caused by a slightly imperfect 3-D tube model, which could degrade the performance especially for the `Center' and `Side' locations, since the aberrations are strongest in these locations. Still, despite these possible imperfections, our model-based method performs mostly comparibly to a sensorless AO exhaustive search, without requiring any feedback measurements.

\section{Conclusion}
We have developed a model-based method for computing wavefront corrections for imaging inside transparent, cylindrical structures of hundreds of micrometers in diameter. With our method, we achieved a 10-fold increase in image contrast for two-photon excitation fluorescence imaging of beads inside a glass tube.

Aside from an initial rotation alignment, which was done both for our method and for sensorless AO, our method does not require any feedback, guide stars or wavefront sensing measurements. Therefore, there is very little photobleaching beforehand, leaving all of the photon budget for imaging.

Our results are on par with an exhaustive-search AO solution, demonstrating that it is possible to model a large 3-D structure with sufficient precision to compute accurate wavefront corrections using \emph{a priori} information only. This finding is particularly relevant since in the case of large aberrations (over 10 radians RMS wavefront error in this case) most efficient AO approaches are not guaranteed to converge to a correct solution. Moreover,  whereas existing methods require additional measurements to extend the imaging range beyond a single isoplanatic patch, our method can easily compute additional phase correction patterns to extend the range, without any additional measurements.

We think our model-based approach to aberration correction could be a great tool in the study of lumen-based organ-on-a-chip systems, as it enables high-resolution imaging inside such type of samples without sacrificing the photon-budget.

\begin{backmatter}
\bmsection{Funding}     
European Research Council under the European Union's Horizon 2020 Programme / ERC Grant Agreement $\text{n}^\circ$ [678919] and the University of Twente.
\bmsection{Acknowledgements} We'd like to thank Gerwin Osnabrugge, Harish Sasikumar, Bahareh Mastiani and Abhilash Thendiyammal for their contributions in constructing the microscope and writing corresponding software.
\bmsection{Disclosures} The authors declare no conflicts of interest.
\bmsection{Data Availability} The ray-tracing code to compute correction patterns is freely available as open source software \cite{IvoVellekoopRaylearn}. The measurement data is available from \cite{coxDataModelbasedAberration2023}.
\end{backmatter}


\begin{thebibliography}{10}
\newcommand{\enquote}[1]{``#1''}

\bibitem{virumbrales-munozMicrofluidicLumenbasedSystems2020}
M.~{Virumbrales-Mu{\~n}oz}, J.~M.~Ayuso, M.~M.~Gong, \emph{et~al.},
  \enquote{Microfluidic lumen-based systems for advancing tubular organ
  modeling,} {\protect\JournalTitle{Chemical Society Reviews}} \textbf{49},
  6402--6442 (2020).

\bibitem{bischelImportanceBeingLumen2014}
L.~L. Bischel, K.~E. Sung, J.~A. {Jim{\'e}nez-Torres}, \emph{et~al.},
  \enquote{The importance of being a lumen,} {\protect\JournalTitle{The FASEB
  Journal}} \textbf{28}, 4583--4590 (2014).

\bibitem{lowOrgansonchipsNextDecade2021}
L.~A. Low, C.~Mummery, B.~R. Berridge, \emph{et~al.}, \enquote{Organs-on-chips:
  Into the next decade,} {\protect\JournalTitle{Nature Reviews Drug Discovery}}
  \textbf{20}, 345--361 (2021).

\bibitem{schwertnerCharacterizingSpecimenInduced2004}
M.~Schwertner, M.~J. Booth, and T.~Wilson, \enquote{Characterizing specimen
  induced aberrations for high {{NA}} adaptive optical microscopy,}
  {\protect\JournalTitle{Optics Express}} \textbf{12}, 6540--6552 (2004).

\bibitem{boothAdaptiveOpticsMicroscopy2007}
M.~J. Booth, \enquote{Adaptive optics in microscopy,}
  {\protect\JournalTitle{Philosophical Transactions of the Royal Society A:
  Mathematical, Physical and Engineering Sciences}} \textbf{365}, 2829--2843
  (2007).

\bibitem{kubbyAdaptiveOpticsBiological2013}
J.~A. Kubby, \emph{Adaptive {{Optics}} for {{Biological Imaging}}} ({CRC
  Press}, 2013).

\bibitem{taoAdaptiveOpticsConfocal2011}
X.~Tao, B.~Fernandez, O.~Azucena, \emph{et~al.}, \enquote{Adaptive optics
  confocal microscopy using direct wavefront sensing,}
  {\protect\JournalTitle{Optics Letters}} \textbf{36}, 1062--1064 (2011).

\bibitem{horstmeyerGuidestarassistedWavefrontshapingMethods2015}
R.~Horstmeyer, H.~Ruan, and C.~Yang, \enquote{Guidestar-assisted
  wavefront-shaping methods for focusing light into biological tissue,}
  {\protect\JournalTitle{Nature Photonics}} \textbf{9}, 563--571 (2015).

\bibitem{vellekoopDigitalOpticalPhase2012}
I.~M. Vellekoop, M.~Cui, and C.~Yang, \enquote{Digital optical phase
  conjugation of fluorescence in turbid tissue,} {\protect\JournalTitle{Applied
  Physics Letters}} \textbf{101}, 081108 (2012).

\bibitem{debarreImagebasedAdaptiveOptics2009}
D.~D{\'e}barre, E.~J. Botcherby, T.~Watanabe, \emph{et~al.},
  \enquote{Image-based adaptive optics for two-photon microscopy,}
  {\protect\JournalTitle{Optics Letters}} \textbf{34}, 2495--2497 (2009).

\bibitem{siemonsRobustAdaptiveOptics2021}
M.~E. Siemons, N.~A.~K. Hanemaaijer, M.~H.~P. Kole, and L.~C. Kapitein,
  \enquote{Robust adaptive optics for localization microscopy deep in complex
  tissue,} {\protect\JournalTitle{Nature Communications}} \textbf{12}, 3407
  (2021).

\bibitem{vellekoopFocusingCoherentLight2007a}
I.~M. Vellekoop and A.~P. Mosk, \enquote{Focusing coherent light through opaque
  strongly scattering media,} {\protect\JournalTitle{Optics Letters}}
  \textbf{32}, 2309--2311 (2007).

\bibitem{kubbyWavefrontShapingBiomedical2019}
J.~Kubby, S.~Gigan, and M.~Cui, \emph{Wavefront {{Shaping}} for {{Biomedical
  Imaging}}} ({Cambridge University Press}, 2019).

\bibitem{mastianiWavefrontShapingForward2022}
B.~Mastiani, G.~Osnabrugge, and I.~M. Vellekoop, \enquote{Wavefront shaping for
  forward scattering,} {\protect\JournalTitle{Optics Express}} \textbf{30},
  37436--37445 (2022).

\bibitem{salterExploringDepthRange2014}
P.~S. Salter, M.~Baum, I.~Alexeev, \emph{et~al.}, \enquote{Exploring the depth
  range for three-dimensional laser machining with aberration correction,}
  {\protect\JournalTitle{Optics Express}} \textbf{22}, 17644--17656 (2014).

\bibitem{matsumotoCorrectionDepthinducedSpherical2015}
N.~Matsumoto, T.~Inoue, A.~Matsumoto, and S.~Okazaki, \enquote{Correction of
  depth-induced spherical aberration for deep observation using two-photon
  excitation fluorescence microscopy with spatial light modulator,}
  {\protect\JournalTitle{Biomedical Optics Express}} \textbf{6}, 2575 (2015).

\bibitem{turagaAberrationsTheirCorrection2013}
D.~Turaga and T.~E. Holy, \enquote{Aberrations and their correction in
  light-sheet microscopy: A low-dimensional parametrization,}
  {\protect\JournalTitle{Biomedical Optics Express}} \textbf{4}, 1654--1661
  (2013).

\bibitem{matsumotoAberrationCorrectionConsidering2018}
N.~Matsumoto, A.~Konno, T.~Inoue, and S.~Okazaki, \enquote{Aberration
  correction considering curved sample surface shape for non-contact two-photon
  excitation microscopy with spatial light modulator,}
  {\protect\JournalTitle{Scientific Reports}} \textbf{8}, 9252 (2018).

\bibitem{yamaguchiAdaptiveOpticalTwoPhoton2021a}
K.~Yamaguchi, K.~Otomo, Y.~Kozawa, \emph{et~al.}, \enquote{Adaptive {{Optical
  Two-Photon Microscopy}} for {{Surface-Profiled Living Biological
  Specimens}},} {\protect\JournalTitle{ACS Omega}} \textbf{6}, 438--447 (2021).

\bibitem{thendiyammalModelbasedWavefrontShaping2020}
A.~Thendiyammal, G.~Osnabrugge, T.~Knop, and I.~M. Vellekoop,
  \enquote{Model-based wavefront shaping microscopy,}
  {\protect\JournalTitle{Optics Letters}} \textbf{45}, 5101--5104 (2020).

\bibitem{DatasheetsDownloadsOptical}
\enquote{Datasheets and downloads for optical glass | {{SCHOTT}},}
  https://www.schott.com/en-gb/products/optical-glass-p1000267/downloads.

\bibitem{tkaczykVectorialLawsRefraction2012}
E.~R. Tkaczyk, \enquote{Vectorial laws of refraction and reflection using the
  cross product and dot product,} {\protect\JournalTitle{Optics Letters}}
  \textbf{37}, 972--974 (2012).

\bibitem{IvoVellekoopRaylearn}
\enquote{Raylearn - {{A}} ray tracer that keeps track op optical pathlength,}
  https://github.com/IvoVellekoop/raylearn.

\bibitem{reddiConvergenceAdam2018}
S.~J. Reddi, S.~Kale, and S.~Kumar, \enquote{On the {{Convergence}} of {{Adam}}
  and {{Beyond}},} in \emph{International {{Conference}} on {{Learning
  Representations}},}  (2018).

\bibitem{vellekoopUniversalOptimalTransmission2008}
I.~M. Vellekoop and A.~P. Mosk, \enquote{Universal {{Optimal Transmission}} of
  {{Light Through Disordered Materials}},} {\protect\JournalTitle{Physical
  Review Letters}} \textbf{101}, 120601 (2008).

\bibitem{sinefeldAdaptiveOpticsMultiphoton2015}
D.~Sinefeld, H.~P. Paudel, D.~G. Ouzounov, \emph{et~al.}, \enquote{Adaptive
  optics in multiphoton microscopy: Comparison of two, three and four photon
  fluorescence,} {\protect\JournalTitle{Optics Express}} \textbf{23},
  31472--31483 (2015).

\bibitem{brugEfficientCartesianRepresentation1997a}
H.~H. van Brug, \enquote{Efficient {{Cartesian}} representation of {{Zernike}}
  polynomials in computer memory,} in \emph{Fifth {{International Topical
  Meeting}} on {{Education}} and {{Training}} in {{Optics}},}  vol. 3190
  ({SPIE}, 1997), pp. 382--392.

\bibitem{coxDataModelbasedAberration2023}
D.~W.~S. Cox, T.~Knop, and I.~M. Vellekoop, \enquote{Data for {{Model-based}}
  aberration corrected microscopy inside a glass tube,
  {{4TU}}.{{ResearchData}},
  {{DOI}}:10.4121/118c6472-dfc4-419b-ba0f-5d2baba77748,}  (2023).

\end{thebibliography}



\end{document}